# A novel approach to the classification of terrestrial drainage networks based on deep learning and preliminary results on Solar System bodies

Carlo Donadio[a] *, Massimo Brescia[b], Alessia Riccardo[a]

Giuseppe Angora[c,], Michele Delli Veneri[d], Giuseppe Riccio[b]

[a] Department of Earth Sciences, Environment and Resources, University of Naples Federico II, 21 Via Vicinale Cupa Cintia, I-80126 Napoli, Italy

[b] INAF - Astronomical Observatory of Capodimonte, 16 Salita Moiariello, I-80131 Napoli, Italy

[c] Department of Physics and Earth Science, University of Ferrara, 1 Via Saragat, I-44122 Ferrara, Italy

[d] Department of Information Technology and Electrical Engineering, University of Naples Federico II, 21 Via Claudio, I-80125, Napoli, Italy

* corresponding author

E-mail: carlo.donadio@unina.it

## Abstract

Several approaches were proposed to describe the geomorphology of drainage networks and the abiotic/biotic factors determining their morphology. There is an intrinsic complexity of the explicit qualification of the morphological variations in response to various types of control factors and the difficulty of expressing the cause-effect links. Traditional methods of drainage network classification are based on the manual extraction of key characteristics, then applied as pattern recognition schemes. These approaches, however, have low predictive and uniform ability. We present a different approach, based on the data-driven supervised learning by images, extended also to extraterrestrial cases. With deep learning models, the extraction and classification phase is integrated within a more objective, analytical, and automatic framework. Despite the initial difficulties, due to the small number of training images available, and the similarity between the different shapes of the drainage samples, we obtained successful results, concluding that deep learning is a valid way for data exploration in geomorphology and related fields.

## 1. Introduction

The hydrographic networks of the Earth represent morphologies of the continental landscape characterized by variable dimensions from hundreds of meters to thousands of kilometers and complex geometry, derived by the mutual interaction of numerous physical, biotic and anthropic factors evolving in space and time due to a non-linear physics [1,2]. The three-dimensional configuration of a watercourse is generally connected to the tectonic-climatic formation conditions and the evolving morpho-climatic system. The recurrence and ubiquity of some networks in different continental areas also depend on regional macroscale aspects, such as geological outcrops and substrate, gradient, climate, type, and extent of the plant cover. At the local microscale, rock fracturing degree, layers attitude, erosive-depositional processes extent, and resistance to erosion are also involved in the modeling process [3,4,5].

However, similar boundary conditions in different regions do not always lead to the same hydrographic patterns modeling: there are phenomena of superimposition of the most recent networks on inherited landscapes, modeled in tectonic and climatic contexts of the past, very different from the current ones. On the contrary, in very different climatic and geological-structural contexts, characterized by active tectonics or volcanism, similar patterns are often observed, due to the morphological convergence and morpho-selection phenomena. These considerations also apply to geometries present in environments very different from terrestrial ones, such as in regions of some Solar System planets and satellites, *e.g.*, Mars and Titan [6,7,8,9].

As known, since the beginning of their history, about 4.5 billion years ago, the Earth and Mars planets shared the phenomenon of the modeling of their surface by watercourses, although today, unlike our planet, the surface of Mars appears completely arid. Even the largest moon in the Saturnian system, Titan, resembles the Earth, being the only planetary body in the Solar System characterized by rivers still flowing today, although in the case of Titan they are powered by liquid methane. Despite this, recently was highlighted how the evolutionary history of Titan is unknown ([10] and references therein). By comparing the various drainage systems present in the two planets and the satellite of Saturn, a greater similarity between those of Titan and Mars was found. In particular, unlike the Earth, in the other two cases the lack of telluric phenomena in the recent past, together with the absence of movements of the tectonic plates, avoided the succession of deviations and anomalies of the river paths, which instead have controlled the modeling of many drainage patterns on the Earth. This means that on the surface of both Mars and Titan, mainly mechanical erosion processes acted. A dense network of large basins was observed on Mars, due to the frequent impact of meteorite bodies and asteroids, which shaped the surface and also conditioned the course of rivers. Recently, this phenomenon was confirmed on Titan by standardizing the resolution of the maps available for the three bodies [10]. This favored the possibility of carrying out a comparative analysis of drainage patterns on different bodies of the Solar System, classifying them directly from the images, fully combining the experience acquired on Earth and the effectiveness of recent data-driven methodologies.

The current classification of hydrographic networks is based on two-dimensional geometry [11]: by eliminating the slope, lithological and climatic factors, the geometric configurations are minimized with a subjective qualitative approach, coming from the observer's experience. For the classification, the geomorphologist focuses on (i) the overall drainage network symmetry and on (ii) the pattern dimensional scaling. The former is rarely present in these natural systems, while the latter is proven mostly for limited portions (sub-basins), rather than for the entire river network. This depends on the difficulty of the geomorphology physics interpretation [12]. Therefore, we tend to identify a series of main and recurrent dimensionless patterns and their derivations (Fig. 1), mostly dependent on lithological and morphological factors [13,14,15,16], independent or not from the regional context.

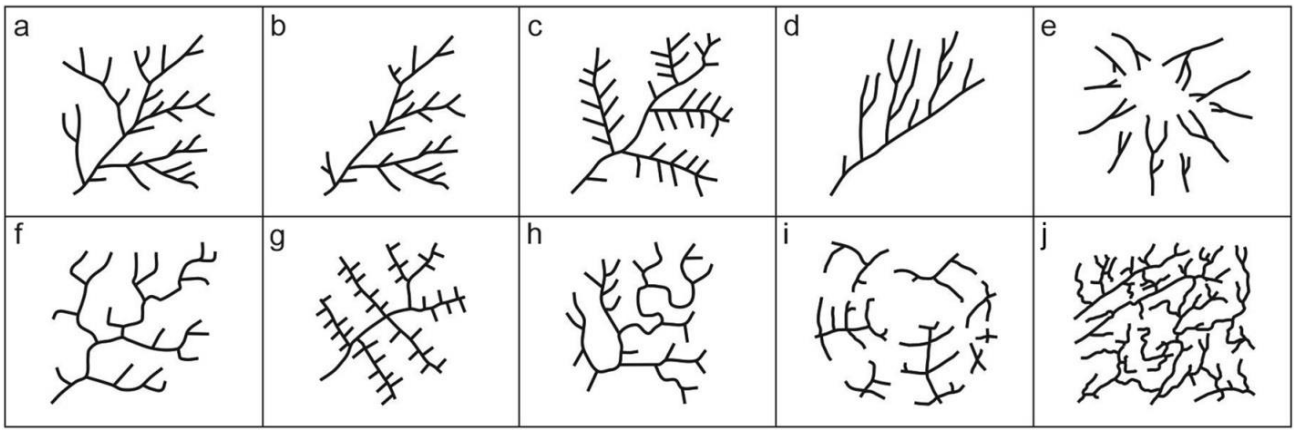

**Fig. 1** - Main patterns of drainage networks: a, dendritic; b, sub-dendritic; c, pinnate; d, parallel; e, radial; f, rectangular; g, trellis; h, angular; I, annular; j, contorted (after [15], [16]). Hereinafter a-c patterns are related to dendritic forms (D), d-l to non-dendritic ones (ND).

The geomorphic-quantitative analysis in river geomorphology [17, 18] provides dimensional parameters and morphometric indices referring to the river length, basin area, hierarchical order, sinuosity, bifurcation, drainage density, etc., useful to compare different networks. Recently, the two-dimensional fractal analysis of some patterns featured by self-similarity [19], together with other metrics, such as the hierarchical order, the sinuosity, and the density of drainage, was proved as a valid tool to discriminate primary (tectonic) processes from secondary ones (erosion), through the interpretation of the fractal dimension. As the latter increases, the effect linked to erosion processes compared to tectonic ones, increases, hence the irregularity of pattern too [20,21,22,23,24,25].

In short, the hydrographic networks appear formed by a set of minor networks of the tributaries of the mainstem, characterized by a different geometry. It is known that the human classification of bi- and tridimensional objects, natural or not, is based on complex and little understood brain perception models, which are also used in Robotics [26]. The apophenia or astronomical "pareidolia", that is the ability to see forms known or not in clouds, mountains, streams, stars, or planets, is one of the most extraordinary illusions of the human mind which permeates countless anthropic and scientific disciplines, such as religion, mythology, art, astronomy, and geology [27]. In such a context, the morphology of drainage patterns is perceived and described based on the prevalent configuration attributed to a single dominant pattern. However, this classification is unsatisfactory and imprecise.

In the analysis of many complex natural systems, e.g. in Astrophysics, Earth and Planetary Sciences [28,29,30,31,32,33,34], the Machine Learning (ML) paradigms allowed a rapid and growing diffusion of Artificial Intelligence (AI) in all sectors of anthropic and scientific activities. The real challenge of AI is that this sophisticated methodology proved to be decisive for computer tasks apparently easy for people, but difficult to describe formally and extremely time-consuming. Among these, there are problems that we solve intuitively and automatically such as recognizing words, faces, or structures within images [35,36,37,38].

The complexity of the human intelligence emulation of natural phenomena by data-driven systems highlights the need to acquire knowledge, extracting conceptual schemes and complex data correlations directly from the primary source, which is the observed data of the studied phenomenon [39,40,41].

The Deep Learning (DL) models incorporate both the characterization of the parameter space and the classification system, becoming particularly suitable in use cases where: (i) the expert user can recognize the structures within images, but at the cost of an impractical and long process, often affected by interpretation controversies; (ii) it is extremely complex to extract the information potentially contained in the images. This implies the intrinsic difficulty of applying traditional analytical methods or classic ML models, particularly sensitive to the negative effects induced by incomplete, noisy, and/or insufficient data.

The scientific problem of an unbiased classification of drainage networks is characterized by the two mentioned aspects, approached through DL in our research. Different from other methods, by adding data, the DL model reinforces the learning power based on its acquired experience, which is used to generalize the ability to classify new samples, thus improving the accuracy and reliability of the classification of new drainage patterns. Such method, directly applied to images, has the real potentiality to define a new, more objective classification of these complex natural elements, characterized by an irregular and asymmetrical geometry, thus contributing to a better characterization and analysis of terrestrial and extraterrestrial examples.

In such a scenario, the primary objective of the present work is to classify the drainage networks in an unambiguous, reliable, and as automated as possible way. Unambiguous means guaranteeing the utmost accuracy in assigning the right class, in accordance, but also on a complementary base, with the broadest and most objective consensus provided by the community of experts in the field.

In order to reach this consensus, this work intends to promote the "River Zoo" survey initiative, aimed primarily at involving the entire scientific community interested in the field and inspired by the well-known category of "citizen science projects". The idea behind such an initiative is mostly to solve the problem of the current lack of drainage network samples useful for facing the exercise of multi-class classification in a statistically consistent and balanced way. Therefore, the multiple participation in the expert survey would be able to guarantee a more reliable assignment of the class to each sample, improving the quality and ensuring an incremental strengthening of the training set for DL models.

The DL approach is also able to guarantee the repeatability, coherence, and consistency of classification, by maximizing the incremental acquisition of experience (incremental learning), thus ensuring the application of same and consolidated criteria to other drainage network samples over time.

Furthermore, another significant aspect of the presented method is to exploit the aseptic and complete information deriving directly from the analysis of the images, avoiding the use of potentially biased, incomplete, and ambiguous derived information, i.e., traditionally extrapolated from processed physical and environmental parameters. Finally, the method is intrinsically automated, thus able to minimize the human intervention in the classification process, relegating it to an a posteriori analysis and the scientific exploitation of the results obtained.

Certainly, the long-term goal of the project, for which this work is a fundamental premise, is the multi-class classification using at least the taxonomy highlighted in Figure 1. However, at the moment, the intrinsic complexity of distinguishing between sub-classes due to the aforementioned ambiguity in the morphology of the patterns and the subjectivity of the attribution of the class, does not allow to have a quantity and quality of examples for each sub-class sufficient to allow the multi-class experiment to be carried out. As it is well known, the supervised paradigm of deep learning requires an adequate number of known examples for each sub-class and requires the fairest possible balance between the quantities of examples for each sub-class. Without this knowledge base, any data-driven method of classification would suffer from underfitting for the under-sampled classes and overfitting for the over-sampled ones. For these reasons, the present work was focused on testing and validating the proposed data-driven method on the reduced two-class problem to distinguish between two families of drainage network sub-classes named, respectively, dendritic and non-dendritic. Moreover, without the experience acquired with this first case, it would be extremely difficult to analyze the results of the more complex multi-class experiment and identify the weaknesses of the method, disentangling the different contributions to the multi-classification error between data-induced errors from those induced by the DL models.

This work is organized as follows: section 2 describes the drainage data used and the image extraction procedure; section 3 explains the classification methods; section 4 reports and discusses the results, while section 5 is dedicated to introducing the public survey initiative named River Zoo, which is open to human experts who want to try to classify the proposed patterns as well as to extend the amount of data samples, aiming at optimizing the knowledge base used to improve the DL training capability.

## 2. Data: drainage networks

Both Earth and extraterrestrial drainage network images have been extracted from public literature and dedicated websites (the list of image references is available[1]. For the supervised DL experiments, we assigned class labels to the collected Earth examples by distinguishing between two classes, respectively, dendritic (D), including subtypes sub-dendritic, pinnate and high-relief pinnate, versus not-dendritic (ND), including other subtypes, such as trellis, parallel, rectangular, angular, annular, radial, centripetal, herringbone and barbed. The intrinsic difficulty to provide a sufficient amount of training samples motivated the choice to apply data augmentation techniques to improve the model training capability. This procedure included uniform sizing of all images to 540x540 pixels and their augmentation, obtaining five further samples for each original image, through three rotations of, respectively, 90, 180, and 270 degrees and two flipping operations concerning, respectively, horizontal and vertical axis. A further advantage was to make the trained model invariant to different orientations of drainage networks within images. Furthermore, we vectorized the images, to make the morphology of a network in a simple, but informative and quantitative form and converted them in grayscale, to simplify the extrapolation of the network pattern. Then the images were cleaned to remove residual noise and to re-arrange parts of the reticle partially lost by the conversion. The labeled dataset of 131 Earth samples was then subdivided into three subsets, using, respectively, 70% for training, 10% for training validation, and the last 20% as a blind test set used to estimate the statistical performances.

---

[1] http://dame.oacn.inaf.it/riverzoo_files/DrainagePatternsReferences.txt

## 3. Method: Deep Learning approach

The quality of the performance of classification systems based on the ML paradigms strongly depends on the effectiveness with which information to be acquired is represented. Every single data or pattern of a scientific domain is an element characterized by a set of parameters with which to represent its informative contribution to the analyzed phenomenon. These parameters are named features, to distinguish them from the inner ML model hyper-parameters. The set of features defines and constitutes the so-called parameter space of a problem.

The dependence on the representation of data in a generic knowledge domain is a significant phenomenon in the scientific field, where the ability to recognize particular complex structures within images, often characterized by poor resolution and low signal/noise ratio, is required.

In the last decade, the DL methodology acquired an ever-increasing application in many scientific fields. Its paradigm is based on the ability to automatically extract the representative features of pixel regions distributed in the image, such as arcs, shapes, and variations in contrast and intensity. These features are then supplied as input to any classification model [42,43].

As known, the experiments based on DL are characterized by a long test campaign for the heuristic optimization of its hyper-parameters, highly dependent on the degree of the intrinsic difficulty of the problem, and only partially mitigated by the experience in the use of such methods. Our test campaign led to the final selection of three customized models, respectively, VGGNet+Adam, VGGNet+RF, and AlexNet+Adadelta (Fig. 2). VGGNet and AlexNet models are two Convolutional Neural Networks (CNNs) known in literature [44,45,46]. The three classifiers were Random Forest (RF; [47]) and two variants of gradient descent methods, respectively, Adadelta [48] and ADAM [49]. The two DL model architectures chosen downstream of a test campaign are two CNNs [43] inspired by two typologies known in the literature. These neural networks are inspired by the behavior of the biological model (human brain). Artificial neurons are organized into several layers, hierarchically connected. As in the biological model, the variation in synaptic connections is related to the learning mechanism. During training, these connections between layers (called weights) are adapted through a mechanism of propagation of the signal back and forth in the network. At the end of the training process, the supervised model (i.e., guided by the knowledge of the expected output value for the training samples) establishes a non-linear input-output relationship, coded by the resulting weight matrix.

The CNN model represents one of the most widespread supervised DL methods, whose peculiarity is the presence of a set of so-called receptor fields that identify the synaptic activity of neurons. A receptor field can be represented by a matrix (also called a kernel or filter) that connects two consecutive layers through a convolution operation. Similar to the mechanism behind the adaptation of weights in a classic neural network, the kernels are modified during training. The important prerogative is the ability of a CNN to automatically extract the representative features of pixel regions distributed in the image, such as arcs, shapes, structures of various forms, variations in contrast and intensity. Then, these features are coded in a vector at the end of the sequence of internal layers, supplied as input to any classification model, which assigns the category to each structure identified in the starting image. The basic idea of a CNN is the multi-layer mechanism in which convolution and synthesis (pooling) levels of the propagated information alternate. This stratification process usually is made up of dozens of layers, in each of which the convolution level acts as a filter, thus emphasizing or suppressing different features of the image. The pooling level ensures the conservation and propagation of essential information (not redundant) extracted up to that point [41]. Unlike the layers of a classic neural network, in which the neurons of one layer are all connected by weights with each neuron of the next layer, in a CNN these connections are "scattered", i.e., limited to a reduced fraction. This reduces the number and complexity of operations, the amount of memory required, and therefore the computing time. The output of each layer consists of a set of sub-images, smaller in size than the original image, called "feature maps".

Another useful operation, even if optional, in the propagation process, is the so-called "dropout", which is a series of random interruptions between the neural connections, to avoid excessive "specialization" of the network to the images provided in phase training and keeping it more "flexible", that is, able to extrapolate the recognition of structures similar but different from those of the training set, a problem known as "training overfitting". In the tests carried out in this work, various dropout percentages were used in some cases, to avoid the occurrence of this problem, together with the use of the common early stopping training method [49].

In the case of classification problems, the final output of a CNN + classifier model is represented in the form of a probability matrix, in which each input sample (image) has associated a probability of belonging to each of the classes of the problem.

The final attribution of the class takes place through the classic "softmax" function, which normalizes an array of values to a probability distribution [50]. The adaptation of the weights of the network, at each training cycle, is performed to minimize an error function (usually called loss function or cost function). In the case of the models used in this work, this function is the "cross-entropy" [51]:

$$C(y, \overline{y}) = \sum_{i=1}^{N_{classes}} y_i ln\overline{y_i} + (1 - y_i)ln(1 - \overline{y_i}) \qquad (1)$$

where y is the expected (target) class and $\overline{y}$ the output of the model.

Amongst the various classification algorithms used during the test campaign, the so-called optimizers, placed as the last layer of the CNN models, three methods resulted in the best candidates:

1. Random Forest: ML method composed of a set of decision trees, built from subsets of features, randomly chosen, capable of attributing, through a "majority" mechanism, the class of belonging to each structure present in the image provided as input [47];

2. Adadelta: optimization method based on the descending gradient technique, which adapts the learning rate to each iteration, avoiding an excessive reduction of the gradient value of the error function [48];

3. ADAM (Adaptive Moment Estimation): a method similar to Adadelta which, in addition to reducing the gradient through the variable learning rate, introduces a mechanism for updating the weights taking into account the moments (mean and variance) of distribution of gradient function [49].

The first CNN model selected was derived from the VGGNet network [45]. The network consists of a series of layers, in which the alternation between convolution and pooling is performed by considering several consecutive convolution levels before each pooling level. This model was chosen by considering its high classification performances, already verified in other scientific fields, characterized by the high complexity of data [52], like those subject of the present work.

The second CNN model selected is inspired by the AlexNet network [45]. In addition to the succession of layers formed by the alternation of convolution and pooling, its main feature is the final "dense" triple-layer, composed of thousands of neurons fully connected.

In terms of qualitative results evaluation, statistical estimators were selected, directly derived from the classic confusion matrix [53]: (i) the average efficiency (AE), obtained from the sum of the correctly classified samples, weighed concerning the total of the samples of the test set; (ii) purity (P), also known as precision; (iii) completeness (C), also known as recall; (iv) F1-score (F1), obtained as the harmonic mean of purity and completeness. These last three estimators were derived for each class (equation 2),

$$AE = \frac{TP+TN}{TP+FP+TN+FN}, P_{pos} = \frac{TP}{TP+FP}, P_{neg} = \frac{TN}{TN+FN}$$
$$C_{pos} = \frac{TP}{TP+FN}, C_{neg} = \frac{TN}{TN+FP} \quad (2)$$
$$F1_{pos} = 2\frac{P_{pos}C_{pos}}{P_{pos}+C_{pos}}, F1_{neg} = 2\frac{P_{neg}C_{neg}}{P_{neg}+C_{neg}}$$

where TP, TN, FP, and FN are, respectively, True Positives, True Negatives, False Positives, and False Negatives (Fig.2). Completeness and purity are the two most interesting estimators to maximize. The completeness measures the ability to extract a complete set of candidates from a class, while purity estimates the ability to select a pure set of candidates from a single class, thus minimizing the level of contamination induced by classification errors (i.e. presence of false positives and false negatives). Usually, the quality of the classification is based on the best compromise between these two estimators.

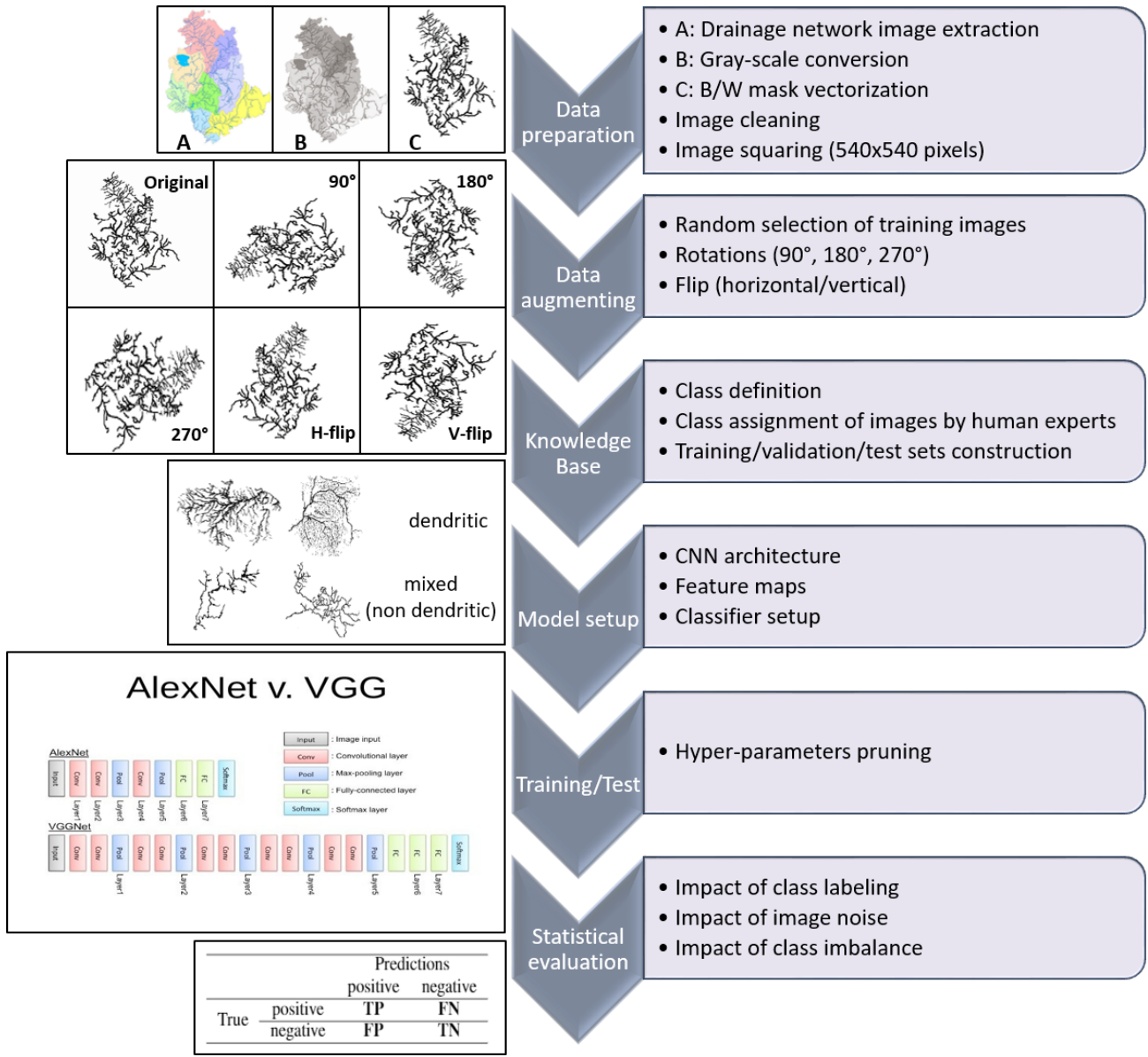

**Fig. 2** - Flowchart of drainage network data processing. On the left, the AlexNet and VGGNet picture, taken from [54]. The bottom left panel shows an example of a confusion matrix for a two-class problem: the rows refer to the expected values, while the columns to the classifier output. True Positive (TP) and True Negative (TN) are the two class quantities of samples correctly classified. FP and FN indicate, respectively, the quantities of False Positives and Negatives.

## 4. Experiments: classification of Earth and extraterrestrial drainage networks

According to what premised in the introduction, three kinds of two-class experiments were performed applying the models previously presented (Tables 1 and 2). First, a qualitative test (hereafter Q-test), to validate the DL approach and to identify the best model. Second, a robustness test (hereafter Rob-test), by replacing the cleaned training and test images used in the Q-test with a noisy pre-cleaning version (see an example in Table 3), to verify the level of robustness to image noise of DL models. Finally, a reliability test (hereafter Rel-test), by replacing the class labeling of the training set, used in the previous two experiments, with a new one in which a portion of training samples was assigned to a different class, according to a more critical analysis by human experts (see the lower part of Table 2).

| **Best Q-test** | **Data type** | **Train set** | | **Test set** | |
|---|---|---|---|---|---|
| | Cleaned data | D: 45 | ND: 58 | D: 13 | ND: 15 |
| | | **Statistics on test set** | | | |
| **Model** | **Class** | **P** | **C** | **F1** | **UC** |
| VGGNet+Adadelta | D | 100 | 100 | 100 | - |
| | ND | 100 | 100 | 100 | - |
| VGGNet+RF | D | 92 | 92 | 92 | Turkey |
| | ND | 93 | 93 | 93 | Devoll |
| AlexNet+Adadelta | D | 100 | 77 | 87 | Tambun Varuna Vishwamitri |
| | ND | 83 | 100 | 91 | - |
| **Best Rob-test** | **Data type** | **Train set** | | **Test set** | |
| | Noisy data | D: 45 | ND: 58 | D: 13 | ND: 15 |
| | | **Statistics on test set** | | | |
| **Model** | **Class** | **P** | **C** | **F1** | **UC** |
| VGGNet+Adadelta | D | 100 | 57 | 73 | Po, Tambun, Varuna |
| | ND | 77 | 100 | 87 | - |
| VGGNet+RF | D | 80 | 62 | 70 | Candelaro, Po, Tambun, Varuna, Vishwamitri |
| | ND | 72 | 87 | 79 | Chari, Devoll |
| AlexNet+Adadelta | D | 92 | 85 | 88 | Tambun, Varuna |
| | ND | 88 | 93 | 90 | Devoll |
| **Best Rel-test** | **Data type** | **Train set** | | **Test set** | |
| | Cleaned data | D: 53 | ND: 50 | D: 13 | ND: 15 |
| | | **Statistics on test set** | | | |
| **Model** | **Class** | **P** | **C** | **F1** | **UC** |
| VGGNet+Adadelta | D | 79 | 85 | 82 | Po, Vishwamitri |
| | ND | 86 | 80 | 83 | Chari, Devoll, Pamba |
| VGGNet+RF | D | 85 | 85 | 85 | Candelaro, Vishwamitri |
| | ND | 87 | 87 | 87 | Chari, Pamba |
| AlexNet+Adadelta | D | 79 | 85 | 82 | Po, Vishwamitri |
| | ND | 86 | 80 | 83 | Chari, Devoll, Pamba |

**Table 1** - Statistical results of the three kinds of classification experiments, performed on the set of 131 drainage networks. P, purity, C, completeness, and F1, F1-score, are in percent. UC reports the list of mismatched samples. D, dendritic, and ND, non-dendritic pattern.

The re-assignment of these labels permitted also a better balancing of the two-class subsets. The latter experiment was particularly suitable to verify the impact of the class labeling, performed by the human experts, on the reliability of the training, reflected on the statistical results on the blind test set. The results of these experiments are reported in Table 1.

In terms of the Q-test, the general validity of the DL approach was confirmed. In particular, the VGGNet+Adadelta model obtained the correct classification of the entire blind test sample. The Rob-test revealed a sensible performance decrease and a strong imbalance among the estimators.

| BLIND TEST SET | | | | | |
|---|---|---|---|---|---|
| **Name** | **Class** | **Cut-out** | **Name** | **Class** | **Cut-out** |
| Anjani and Jhiri, India | D | 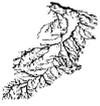 | Olenek, Northern Siberian Russia | ND | 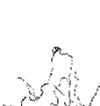 |
| Bunta, Indonesia | D | 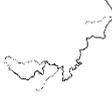 | Ottawa,, Canada | D | 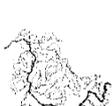 |
| Candelaro, Italy | D | 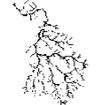 | Pamba India | ND | 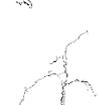 |
| Chalakudy, India | ND | 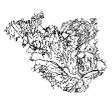 | Pinios, Greece | ND | 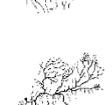 |
| Chari, Central Africa | ND | 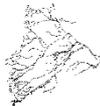 | Po, Italy | D | 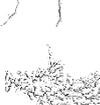 |
| Devoll, Albania | ND | 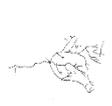 | Sava, Central Europe | ND | 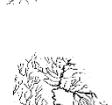 |
| Huang He, China | ND | 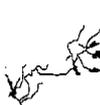 | Snake Northwest America | D | 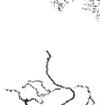 |
| Jordan, Western Asia | ND | 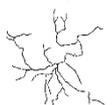 | Tambun, Indonesia | D | 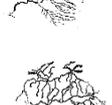 |
| Karuvannur, India | D | 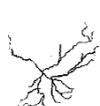 | Tevere, (taly | D | 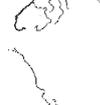 |
| Lena, Eastern Siberia | ND | 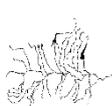 | Turkey, Midwest America | D | 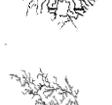 |
| Mekong, Southeast Asia | ND | 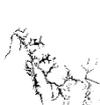 | Varuna, India | D | 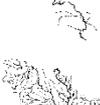 |
| Mississippi, Northern America | D | 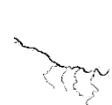 | Vishwamitri, India | D | 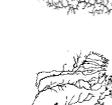 |
| Muvattupuzha, India | ND | 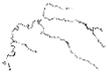 | Volga, Russia | ND | 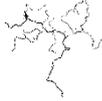 |
| Ob, Western Siberia | ND | 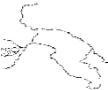 | Yenisey, Russia; Mongolia | ND | 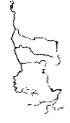 |

**TRAINING SET**

| | | | | | |
|---|---|---|---|---|---|
| AchanKovil, India | [ND] D* | 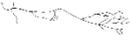 | Pinamula, Indonesia | [D] ND* | 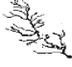 |
| Adda, Italy | [ND] D* | 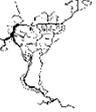 | Loira, France | [D] ND* | 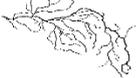 |
| Bahomoleo, Indonesia | [ND] D* | 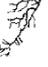 | Sabarmati, India | [ND] D* | 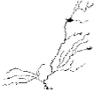 |
| Bharathappuzha, India | [D] ND* | 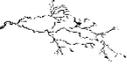 | Simeto, Italy | [D] ND* | 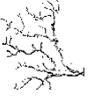 |
| Luni, India | [ND] D* | 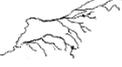 | Toaya, Indonesia | [ND] D* | 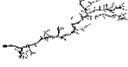 |
| Manimala, India | [ND] D* | 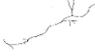 | Tobol, Russia, Kazakhstan | [ND] D* | 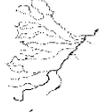 |
| Meuse, Western Europe | [ND] D* | 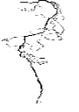 | Volturno, Italy | [ND] D* | 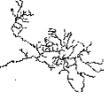 |
| Ombrone, Italy | [ND] D* | 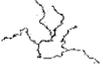 | Yangtze, China | [ND] D* | 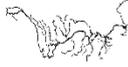 |

**Table 2** - Upper part: list of drainage networks composing the blind test set used in the experiments, where D is dendritic, ND non-dendritic; lower part: list of training drainage networks with the first-class assignment (the class between square brackets), used in the Qualitative test (Q-test) and robustness test (Rob-test) experiments. The class marked with an asterisk indicates a different class assignment as changed in the training set for the reliability test (Rel-test).

| Name | Cut-out | |
|---|---|---|
| | noisy | cleaned |
| Amazon | 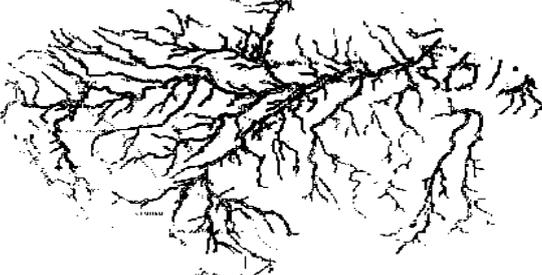 | 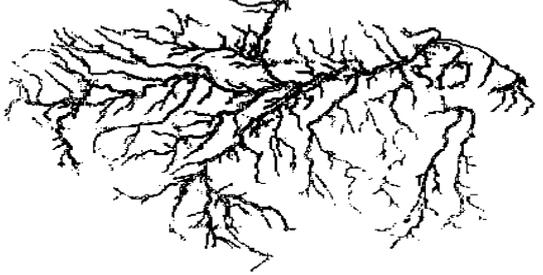 |
| Mackenzie | 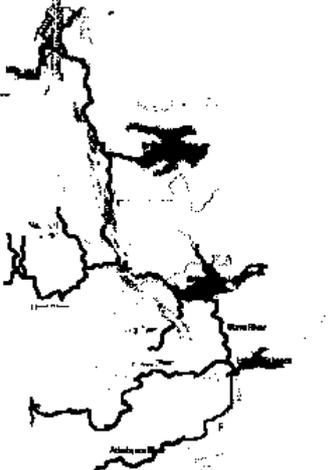 | 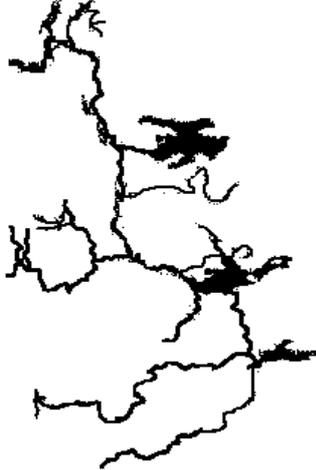 |

**Table 3** - Examples of noisy and cleaned training drainage networks.

This behaviour shows that the feature maps produced by CNNs were sensible to the presence of residual image noise. In practice, the residual image noise increased the already intrinsic difficulty of the problem.

The Rel-test showed a more balanced result among statistical estimators than the Rob-test, but always lower than the Q-test, reflecting the high relevance of the right class labeling. In practice, the Rob-test and Rel-test showed that the DL models were pushed to their classification capability limits. This, although a high contribution to such behavior is probably due to the limited amount of samples available, only partially mitigated by data augmenting. Later, the models trained on Earth samples were applied to extraterrestrial drainage networks (Table 4).

| Name | Class | VGG Adelta [RF] | ANet Adelta | Cut-out | Name | Class | VGG Adelta [RF] | ANet Adelta | Cut-out |
|---|---|---|---|---|---|---|---|---|---|
| Aeolis Planum, Mars | ND | ND [ND] | ND | 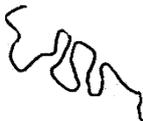 | Osuga Valles, Mars | ND | ND [ND] | ND | 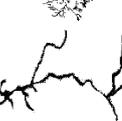 |
| Arabia Quadrangle, Mars | ND | ND [ND] | ND | 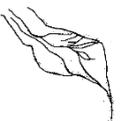 | Huygens Crater (east); Hellas Crater (north), Mars | ND | ND [ND] | ND | 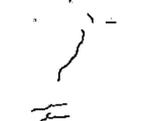 |
| Chryse Planitia, Mars | ND | ND [ND] | ND | 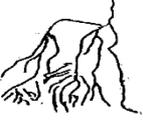 | Paraná Valles, Mars | ND | ND [ND] | ND | 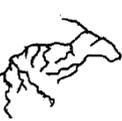 |
| Cusus Valles, Mars | ND | ND [ND] | ND | 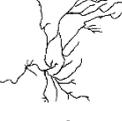 | Efesto Fossae, Mars | ND | ND [ND] | ND | 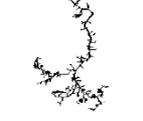 |
| Eberswalde Delta Crater, Mars | D | D [D] | **ND** | 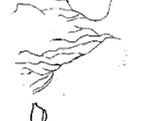 | Schiaparelli Crater, Mars | D | **ND** [**ND**] | **ND** | 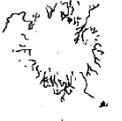 |
| Gale Crater, Mars | ND | ND [ND] | ND | 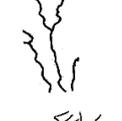 | Terra Cimmeria, Mars | D | **ND** [**ND**] | **ND** | 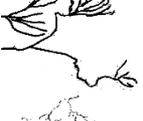 |
| Indus Vallis, Mars | ND | ND [ND] | ND | 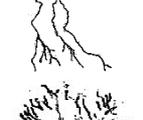 | Terra Cimmeria1, Mars | D | D [D] | D | 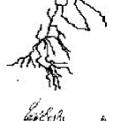 |
| Locras Valles, Mars | D | D [D] | **ND** | 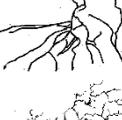 | Titan River2, Titan | ND | ND [ND] | ND | 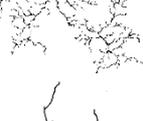 |
| Ma'adim Vallis, Mars | ND | ND [ND] | ND | 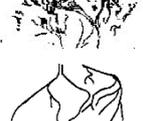 | Titan River, Titan | ND | ND [ND] | ND | 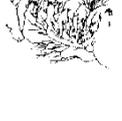 |
| Ma'adim Vallis2, Mars | ND | ND [**D**] | ND | 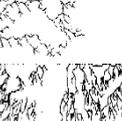 | Tyras Vallis, Mars | ND | ND [ND] | ND | 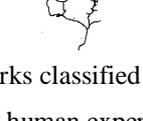 |
| Eberswalde Delta Crater2, Mars | ND | ND [ND] | **ND** | 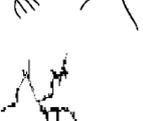 | Warrego Valles, Mars | D | **ND** [**ND**] | **ND** | 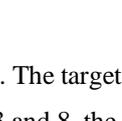 |
| Naktong Vallis, Mars | D | D [D] | **ND** | 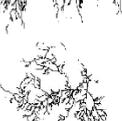 | Warrego Valles2, Mars | D | D [D] | D | 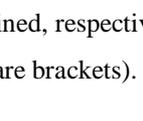 |
| Nirgal Vallis, Mars | ND | ND [ND] | ND | 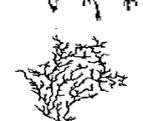 | | | | | |

**Table 4** - Extraterrestrial drainage networks classified by the three models trained on terrestrial samples. The target class (columns 2 and 7) was assigned by human experts: D is dendritic, ND non-dendritic. In columns 3 and 8, the double class value refers to the results obtained, respectively, by the VGG+Adadelta (upper value) and by the VGG+RF models (lower value in square brackets). Mismatches in the class are highlighted in bold.

## 4. Discussion

This research offers a new perspective on tackling the drainage network classification problem. A descriptive approach is usually followed in literature [11,13,14,15,16,17,18,19]. Alternatively, our innovative approach extrapolates the geomorphic network expression directly from images. In this way, it was also possible to verify the degree of divergence among experienced observers in terms of classification, highlighting how much the perception of a given pattern is subjective and based on simplified schemes, morphologically distant from natural phenomena. An optimal classifier, the VggNet+AdaDelta model, resulted from an intensive test campaign. The challenge was to find a right DL architecture, trained on a limited and imbalanced data sample. Despite such limitations, the model showed on average high prediction percentages, ranging between 93% and 100%. Therefore, the DL used for the first time to recognize geomorphological objects can provide a significant contribution to the research field if we consider that, without these techniques, the aforementioned "features" would be produced and evaluated manually, as well as classified with a level of low and subjective predictability.

In terms of lessons learned from experiments, by keeping the best model and labeling unchanged, the random choice of training/test examples has sometimes decreased the performances, revealing the strong impact of training data. This is due to the intrinsic characteristics of some drainage network morphologies and their frequent strong similarity. Furthermore, as the labeling changed, the DL models presented different behaviour. This could be due to the classification statistical criterion, based on the principle of maximum likelihood, whose probability corresponds to the proximity to the target class [55]. Consequently, various image regions are probably at the limit between the two classes, making models very sensible to any class attribution change. Finally, the choice of binary classification may have influenced the performance "fluctuations" as the labeling changed, mostly for the non-dendritic class, composed by a large variety of drainage networks. We underline that the chosen simplistic duality dendritic/non-dendritic was related to families of drainage patterns, not only to two single morphological types, as described in Sec.2. This approach is typical at the beginning of such kind of multi-class experiments based on ML methods, where the refinement of the sub-class recognition can be pursued after having acquired the needed experience and validated the methodology on the "apparently" simplest case. But such refinement requires a consistent training sub-class sample, only partially available at the moment and for which we proposed the public survey River Zoo (see section.5).

However, the results were surprising, considering the quantity and quality of the images, affected above all by problems of uniformity, noise, and homogeneity, unevenness of the monochromatic tone, presence of small holes, imprecise and frayed boundaries of each segment, as well as by the class imbalance and the absence of scientific literature to refer to as a starting point.

Considering the quantitative geomorphic and fractal analyses, subject of traditional works [5,18,19,22,24], our data-driven approach based on the direct application of DL to the drainage pattern images, overcomes intrinsic limits of traditional quantitative methodologies, by contributing at the same time to the improvement of the drainage pattern classification as well as to their correlation with climatic-geomorphological aspects.

The proof of DL efficiency was confirmed by the results obtained on the 25 extraterrestrial drainage networks. This additional preliminary exercise of classifying drainage networks of other Solar System bodies has the primary purpose of analyzing and evaluating the degree of generalization of DL models, also verifying the objective capacity to recognize well-defined categories of drainage networks in different geo-evolutionary contexts, based on learning and the degree of accuracy achieved on terrestrial samples. This will be used in further works, hopefully, thanks to the River Zoo outcomes (see section.5), as a prerequisite for the recognition of anomalies in extra-terrestrial contexts, probably due to the specific morphological and evolutionary features present on other planets.

These patterns, in particular those of Mars and Titan, despite having some morphologies similar to those on Earth, were modeled in very different geological and climatic environments. On Mars, gravity (0.371g) is about 2.65 times lower than Earth (0.981g). In this case, the drainage networks were not modeled due to the lower weight of water and sediments, but from the higher speed of water and the intense surface run-off in the past [56], in a more rarefied, cold, and arid atmosphere. This highlights a correlation between branch angles and climate control, so similar drainage patterns are likely to develop in similar climates [57]. Instead, Titan patterns were modeled in an environment with gravity 0.135g, about 1/7 of Earth, in rocks and fluids composed of frozen hydrocarbons, i.e. methane and ethane in solid and liquid phases, in an atmosphere 1.19 times denser than the terrestrial one, cold and with low viscosity [58]. These represent cases of morphological convergence of erosion shapes [59], suggesting that some patterns could be ubiquitous in many planets and moons of the Solar System.

By predicting the class of extraterrestrial patterns, using models trained only on Earth samples, the results were in agreement with human expert analysis in 88% of cases. Therefore, the DL approach appears quite robust in the prediction, even in cases of different morphologies, and with a high rate of generalization accuracy.

## 5. The River Zoo initiative

As outlined in the previous sections, this work was mainly based on two fundamental aspects, both deriving from the intrinsic multidisciplinary nature of the pursued approach.

First, the absolute novelty of the application of methodologies, already widely used and validated in other scientific fields, to a specific scientific use case, in this case, the geomorphology one. This had the primary purpose of (i) minimizing, in perspective, the subjective human interpretation in the classification of drainage patterns and (ii) performing the pattern recognition directly from images, thus avoiding any bias effect induced by the selection of the parameter space, i.e. the physical and environmental parameters extracted from images, which is at the base of most traditional approaches.

Secondly, due to the novelty of the approach, we had to proceed from the initial level, that is, from the simplest and most general possible case of the two macro-class classification, for instance, dendritic versus non-dendritic, avoiding in the first instance the well-known problem of error propagation in the case of a hierarchical pairwise classification of different sub-classes. This, to validate the methodological approach and to acquire the necessary experience to be subsequently poured into the harder multi-class problem. Naturally, the simultaneous multi-class classification, that is the attribution of a drainage pattern to one of the ten classes, enucleated in Figure 1, in a single experiment, is the most interesting final goal, which our work aims at. However, to pursue this final goal, a statistically consistent and balanced knowledge base is required to properly train the models, in which there should be a sufficient representative sample of all the involved classes.

Precisely for this reason, we launched the River Zoo initiative, which aims at quantitatively enriching the database, both in terms of quantity and variety of examples suitable for training the Deep Learning models, through the direct involvement of the wider community of experts interested in the scientific problem.

Due to the high sensitivity to the subjective class labeling of training samples, we implemented a public survey addressed to interested researchers, called "River Zoo" and available on the Web[2]. Such initiative, inspired by what was done in Astrophysics [60], by exploiting the scientist experience, has the role to provide reliable ground truth for DL-based classification.

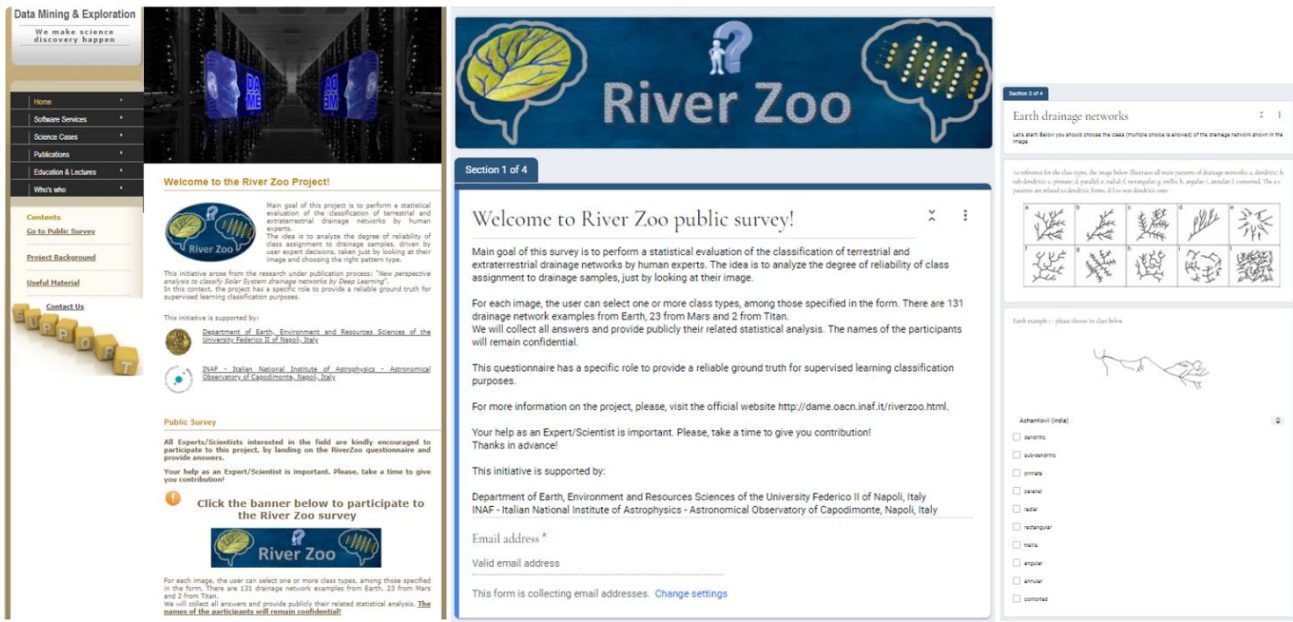

**Fig. 3** - The River Zoo and related survey web pages.

---

[2] http://dame.oacn.inaf.it/riverzoo.html

Main goal of this project is to perform a statistical evaluation of the classification of terrestrial and extraterrestrial drainage networks by human experts.

The idea is to analyze the degree of reliability of class assignment to drainage samples, driven by user expert decisions, taken just by looking at their image and choosing the right pattern type (Fig.3). In this context, the project has a specific role to provide reliable ground truth for supervised learning classification purposes. We expect that the community involved in this field could contribute by suggesting more samples of both terrestrial and extraterrestrial drainage patterns, thus increasing the size of the incremental training data with time.

## 6. Conclusions

The DL approach represents a useful method for the analysis and classification of Earth and extraterrestrial physical geomorphological objects like drainage networks, here applied for the first time.

The study of terrestrial drainage patterns is suitable for the research of geological processes responsible for the formation of a specific morphology and may have significant results even at a multidisciplinary level. Considering that the geological-climatic context and the network shape are mutually interconnected, the analysis of Earth, Mars, and Titan basins assumes an important relevance for the study of paleoclimate and indicates a morphological convergence of different and complex erosion processes, suggesting that some patterns could be ubiquitous in the Solar System. Finally, we are confident that the River Zoo public survey initiative could contribute to improving the knowledge about morphological and geologic-climatic aspects of drainage patterns, besides its potential to improve their classification and to extend the training ground truth.

## Online content

All methods, materials, and data are available upon motivated request to the authors.

**Acknowledgements**

The software for the Deep Learning models was developed within the DAME project [32]. MB acknowledges the INAF PRIN-SKA 2017 program 1.05.01.88.04. This study was supported during the academic year 2017-18 by the International Exchange Program between the University of Naples Federico II (Italy) and Foreign Universities or Research Institutes for Short Time Mobility of Teachers, Scholars, and Researchers. CD stayed as a visiting scholar at the Institute of Urban and Regional Development, Department of Landscape Architecture and Environmental Planning, University of California, Berkeley (USA).

**Author contribution**

CD, conceptualization, writing, supervision; MB, experiments, software design and development, supervision; AR, images preparation, experiments; GA, software development of Deep Learning models and classifiers; MDV, software development for data augmenting; GR, hardware configuration and setup for experiments.

**Competing interests**

The authors declare no competing interests.